# Color Perception: Opening up the Chromaticity Cone

Prashanth Alluvada[*]
Department of Mathematics
Tulane University, New Orleans, LA, USA
Email: prashanthalluvada@gmail.com

**Abstract**

In the XYZ color space, the subset of the tri-stimuli corresponding to spike-type (monochromatic) impingement of energy is the chromaticity cone (CC). Using a family of concentric spheres, we describe a nonlinear transformation over the CC and construct a bijection from the CC onto the (flat) plane. In the process, we open up the CC and view it as a chart on the plane. Because the map is a bijection, the color perception information is preserved (invariant) through the transformation. We call this flat image of the CC, the "chromaticity chart". We mark the isoX, iso-luminance (isoY) and isoZ lines and the geodesic lines on the chart. Further, we add infinity to the chromaticity chart and obtain a one point compactification of the entire chart plus infinity through a stereographic projection. The stereographic projection then lets us to view the entire compactified chromaticity chart plus infinity as an image on the unit sphere. In the XYZ space, ideally, arbitrary heights for the monochromatic energy are admissible, which means that the CC could be indefinitely long. By adding infinity to the chromaticity chart, the entire cone including the case of infinite energy gets mapped onto a compact sphere, the sphere's pole corresponding to the infinite energy. This step maps the entire range for color perception due to spike type energy impingement, including the spikes of infinite heights, onto a compact unit sphere.

1. Introduction

In this article, the chromaticity cone (CC) is opened up as a flat chart. The constructions are made in the XYZ color space. The sections of the CC with X=constant, Y=constant and Z=constant planes, respectively referred as isoX, iso-luminance, isoZ are marked on the chart. The cone's geodesic lines are also marked on the chart. A family of concentric spheres is first superposed over the CC. The center of the sphere coincides with the origin of the XYZ color space. The arc-lengths of the space-curves of intersection (spheres-CC) are computed and used to map the cone onto the plane. The plane of purples is mapped onto the plane using their corresponding space curves of intersection (arcs of great circles). All these maps are bijections.

2. The Metric of the Chromaticity Cone

The color matching functions shown in Fig. 1. are used. If the height of the spike type spectral radiance is 'h' and the wavelength = lambda, the equations for the chromaticity cone are:

$$X = hx(\lambda)$$
$$Y = hy(\lambda) \qquad (1)$$
$$Z = hz(\lambda)$$

then the metric on the chromaticity cone is

$$\begin{aligned} ds^2 &= dX^2 + dY^2 + dZ^2 \\ &= (x(\lambda)^2 + y(\lambda)^2 + z(\lambda)^2)dh^2 + h(x(\lambda)'^2 + y(\lambda)'^2 + z(\lambda)'^2)d\lambda^2 \end{aligned} \qquad (2)$$

The geodesics of the chromaticity cone are the extremals of the following integral (obtained by applying the Euler Lagrange equations and variational calculus for this situation).

$$I = \int_{\lambda_1}^{\lambda_2} ds$$

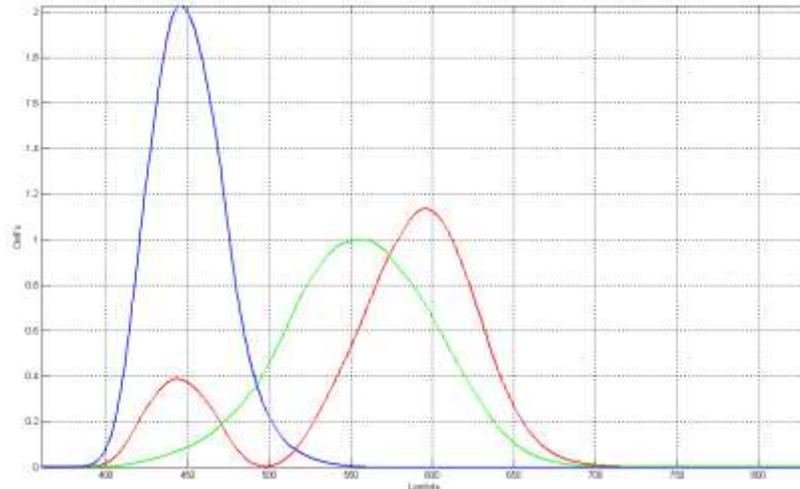

**Fig. 1.** The color matching functions used for the calculations of this article.

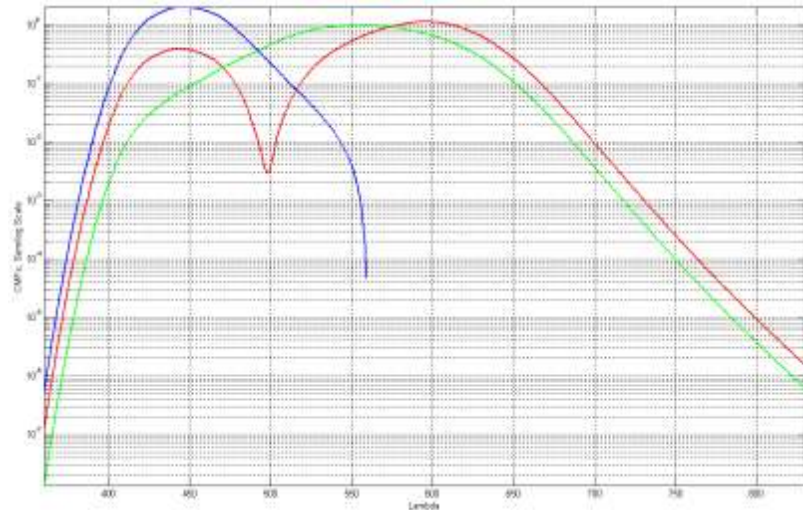

**Fig. 2**. The same cmf's on a semilog scale.

The chromaticity cone is a made up through space curves of the type shown in Fig.3. The space curve is obtained by varying lambda at a constant h. In Fig.4. four space curves are shown each corresponding to a different h, each sweeping the full lambda range. Therefore when the h is made to vary continuously, the space curves (of Fig.3) become continuous and generate the CC. This observation is the key to the opening up of the chromaticity cone, for this helps to develop a bijective correspondence between the cone and the plane. On the other hand, varying the h at a fixed wavelength produces a straight line in the XYZ space, passing through the origin. In this type of variation, the CC is generated as a family of straight lines, all passing through the origin of coordinates. This situation is depicted in Fig.5.

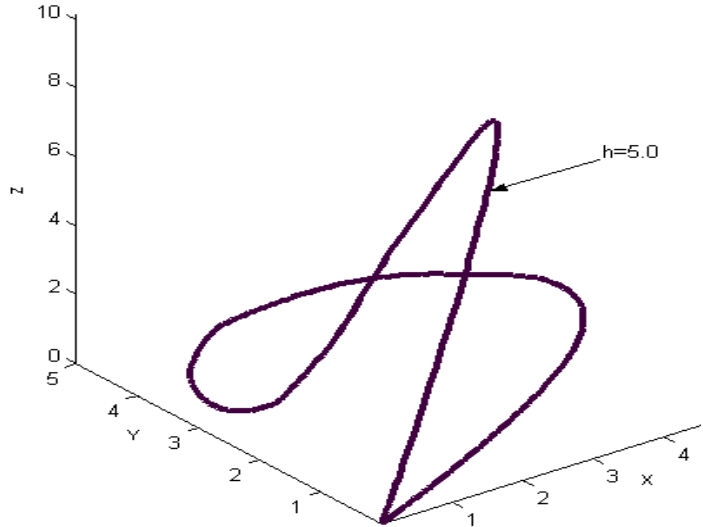

**Fig.3.** For a constant "h" and varying all lambda, the space curve in the XYZ space is shown.

The plane of purples is obtained by combining various proportions of the red and the violet wavelengths (two spikes) and connects the open ends of the chromaticity cone. Upon exhausting all single spikes of all heights, over all wavelengths, the CC is spanned with the ends still open (unconnected). The two open ends (red-end and the violet-end) of the cone are connected using linear combinations of two-spikes (of red and violet) and spans the plane of purples. The cone together with the plane of purples defines the region for color perception for a standard observer (with appropriate assumptions holding). We note here that the XYZ space is a linear transformation away from the RGB space, which carries the actual data from the color matching experiment.

3. A Concentric Family of Sectioning Spheres:

To open up the chromaticity cone, a family of concentric spheres is placed over the XYZ color space, with centers coinciding with the origin of coordinates of the color space. The space curves of intersection of the unit sphere with the CC is computed. Because of the geometric similarity of the straight line, other space curves of sphere-cc intersection may be constructed from the space curve of the unit sphere-CC intersection, by means of a scaling factor composed through the radii of the spheres.

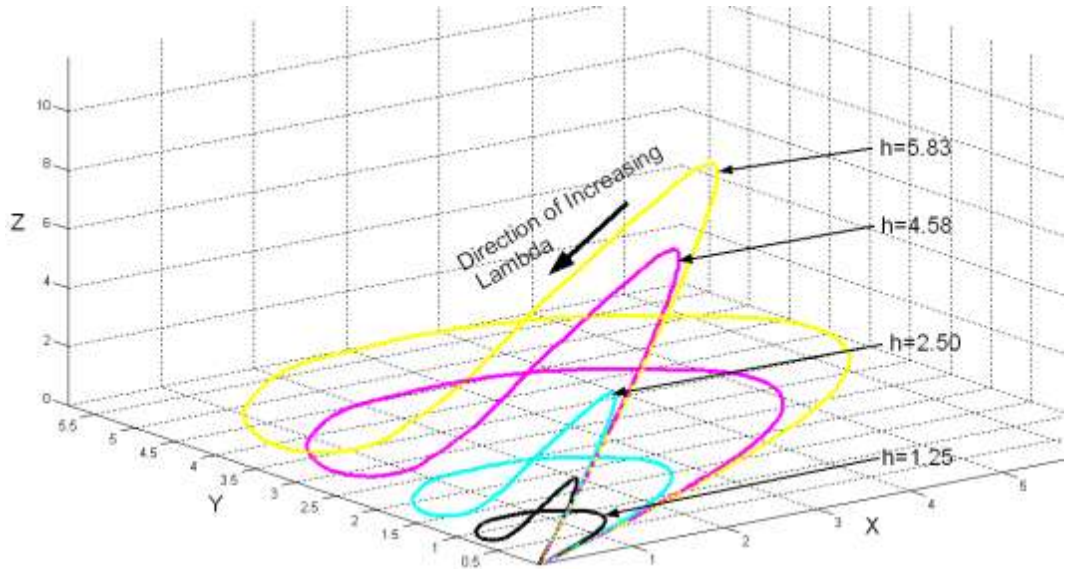

**Fig.4.** It follows from the definitions that space curves of the type shown here make up the chromaticity cone. Each is obtained by keeping the "h" constant and sweeping through the entire range of wavelengths of the visible spectrum. Four different colors are used only to depict the four different space curves.

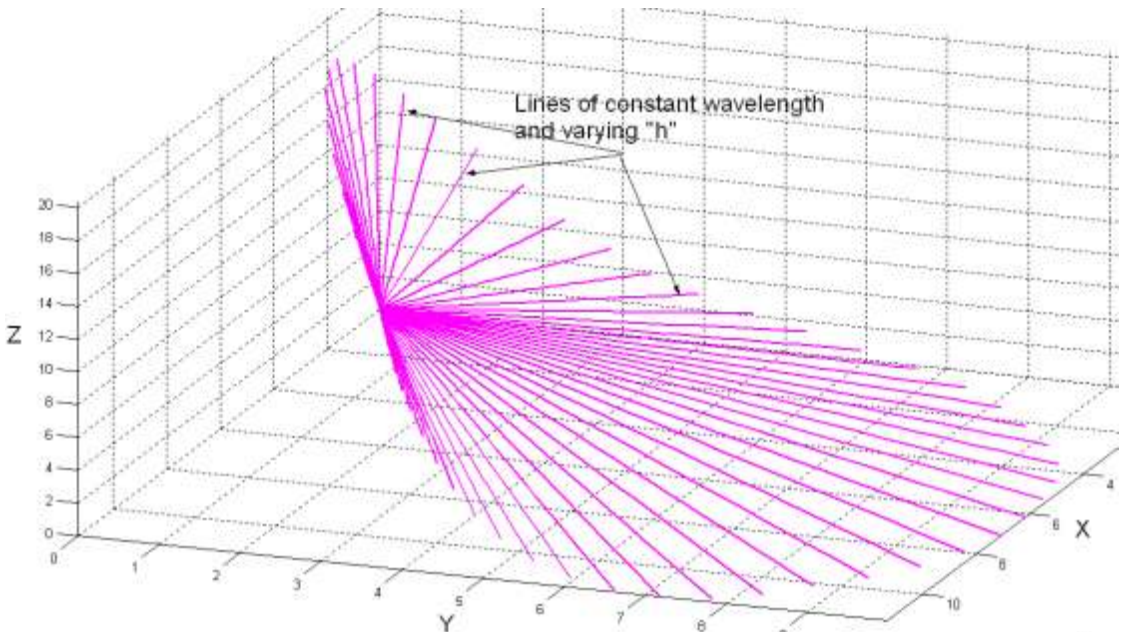

**Fig.5.** Lines of constant lambda and varying "h" (spike height). The CC is spanned as a family of straight lines through the origin.

The arc lengths of the space curves of intersection of the spheres with the chromaticity cone are then computed. A conveniently chosen point on the space curve may be used as a reference point for measuring the arc length. In this article, the intersection of the line [x(end) y(end) z(end)] with the unit sphere is chosen for the purpose. Because arc length is an intrinsic property of the space curve, the length up to the current point is sufficient and necessary for uniquely mapping the point onto the base circle corresponding to the current point on the sphere-CC. This step determines the angle subtended by the image of the current point, at the center of the base circle (see Fig.10 for base circle). This correspondence is a bijection because given the CC and the chromaticity chart, the mapping from the CC to the chart is clear. Now to obtain the point on the CC corresponding to a given point on the chart, we first get the polar coordinates of the point on the chart. Then h-lambda are computed using the following steps: first the lambda corresponding to the theta is obtained from the theta-lambda curve. The curve should be scaled using the radius of the sphere corresponding to the point in question. Then on getting the lambda, the h is computed using the equation,

$$h = \frac{r}{\sqrt{(x(\lambda)^2 + y(\lambda)^2 + z(\lambda)^2)}}$$

where r is the radius of the sphere.

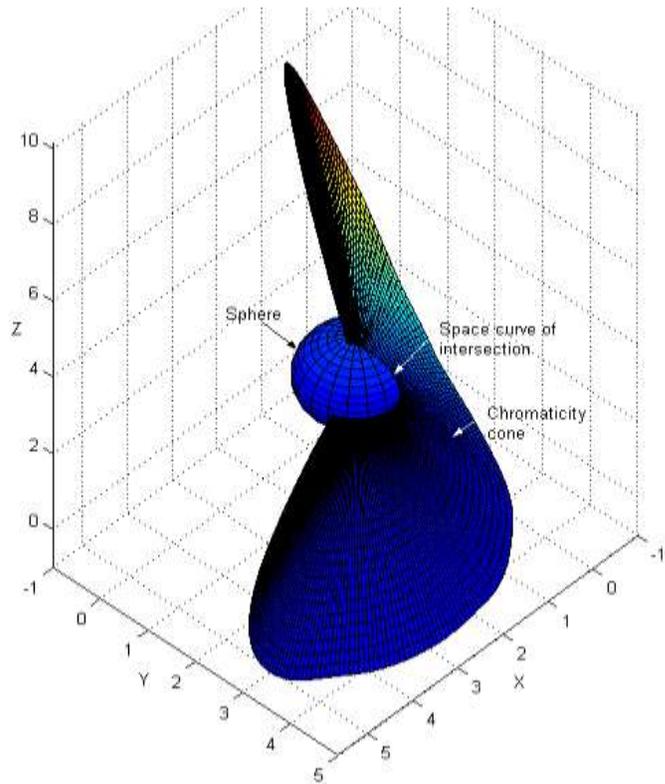

**Fig.6.** The intersection of the chromaticity cone and sphere is shown. The space curve of intersection, labeled in the picture, is aligned along the base circle in such a way that the arc lengths of the space curve determine the subtended angles at the center of the base circle.

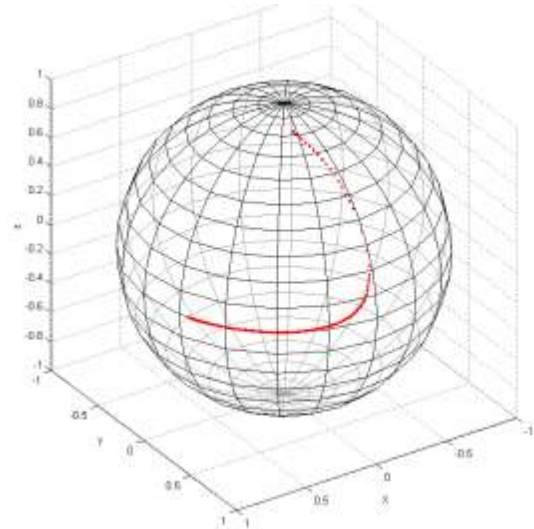

**Fig.7.** The space curve of the sphere-cc intersection is shown on the sphere, removing the cc from the view. This space curve is aligned along the base circle (sphere-XY plane intersection) so that the (progressive) arc lengths (along the curve) determine the angles subtended at the center of the base circle.

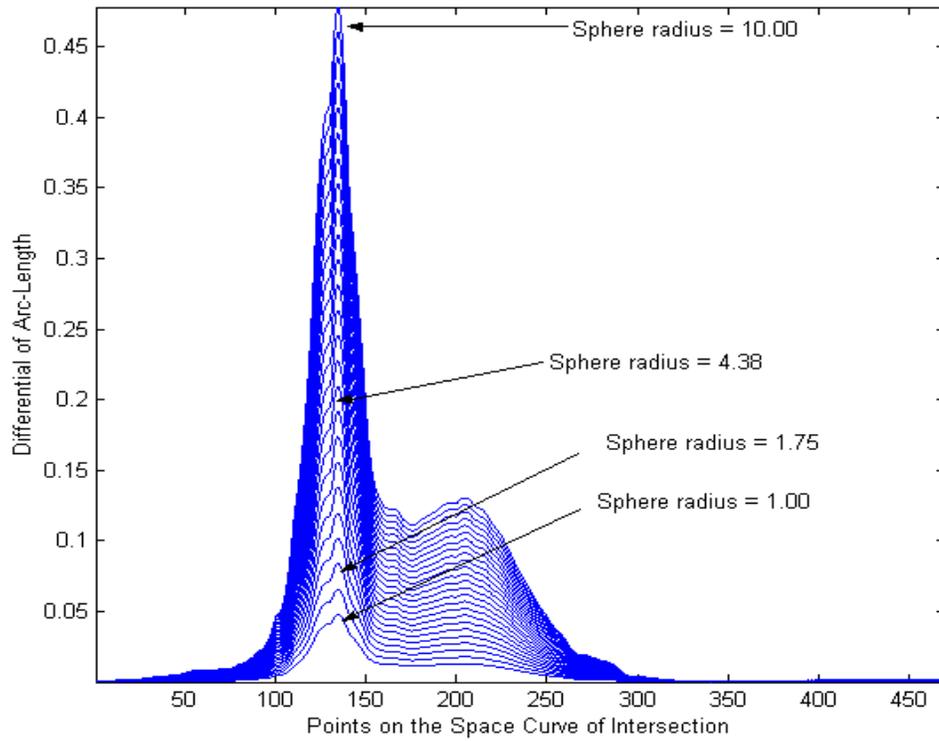

**Fig. 8.** The differential arc lengths of the space curve of the sphere-cc intersection is shown, for spheres of varying radii. The cumulative sum of these differential lengths up to a chosen point gives the arc length up to that point. Curves corresponding to twenty-five spheres sectioning the CC are shown. This metric is at Eqn.(5).

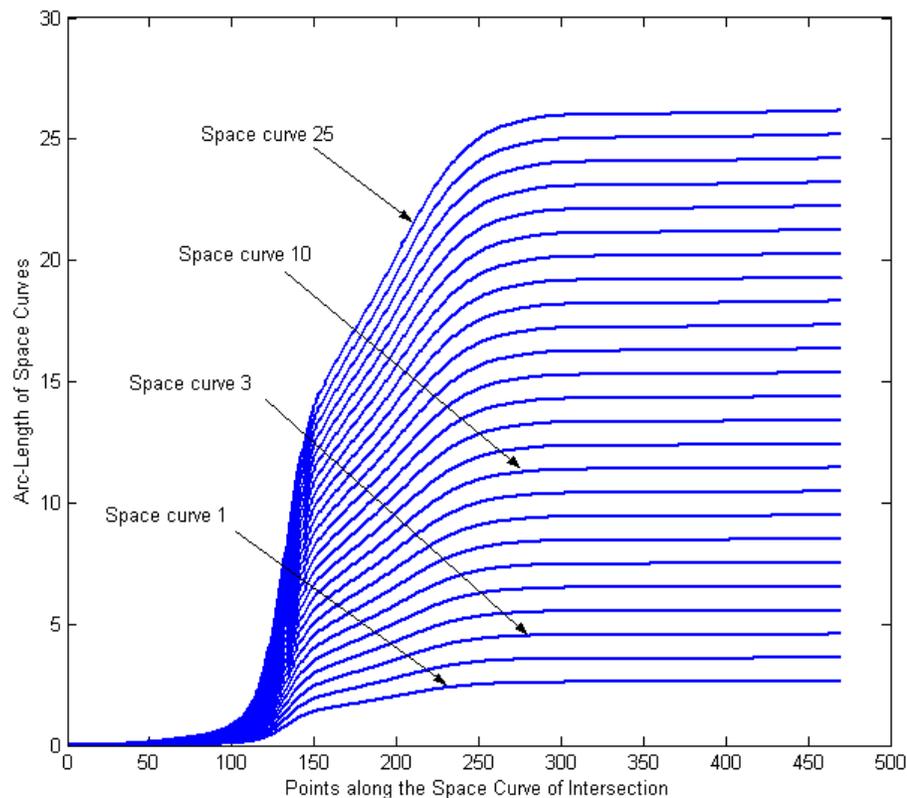

**Fig. 9.** The arc lengths of the space curves of spheres-cc intersections. On the abscissa is the point on the space curve and the ordinate shows the arc length of the space-curve up to that point. These curves are the cumulative lengths of the differential arc lengths shown in Fig. 8

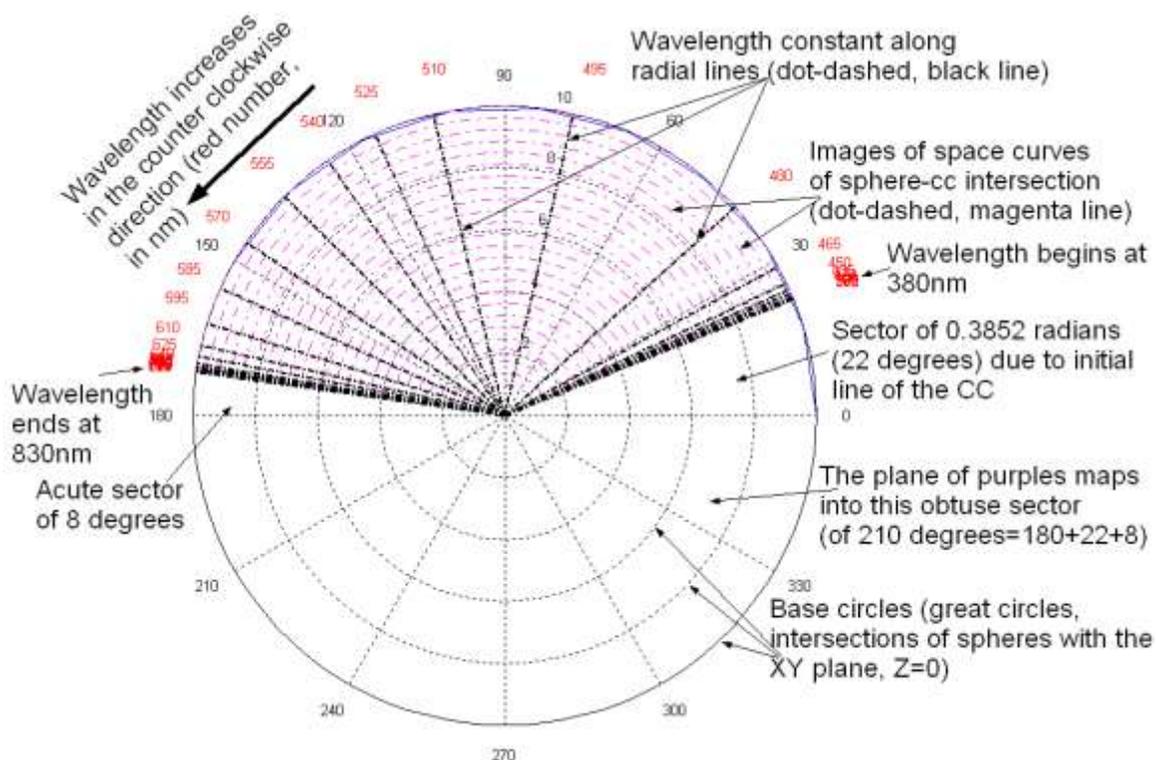

**Fig.10.** Opening up the chromaticity cone using the space curves of intersection results in the chromaticity chart shown above. The wavelength is constant along the radial lines. Its measured counter-clockwise. The concentric arcs of circle perpendicular to radial lines (marked black) are the images of the space curves of the sphere-cc intersection.

To map the plane of purples, onto the plane, the following steps are used:

1. The curve of intersection of plane of purples with the current sphere is considered. It is evident that the curve is an arc of the great circle of the sphere. It is marked as Arc of Great circle 1 in the figure.
2. The curve of intersection of the horizontal plane with the same sphere is a great circle. The segment is marked arc of great circle 2, is chosen as the image of the former segment. The points of these arcs are mapped bijectively and the correspondence is made using the arcs of great circles, as shown in the figure.

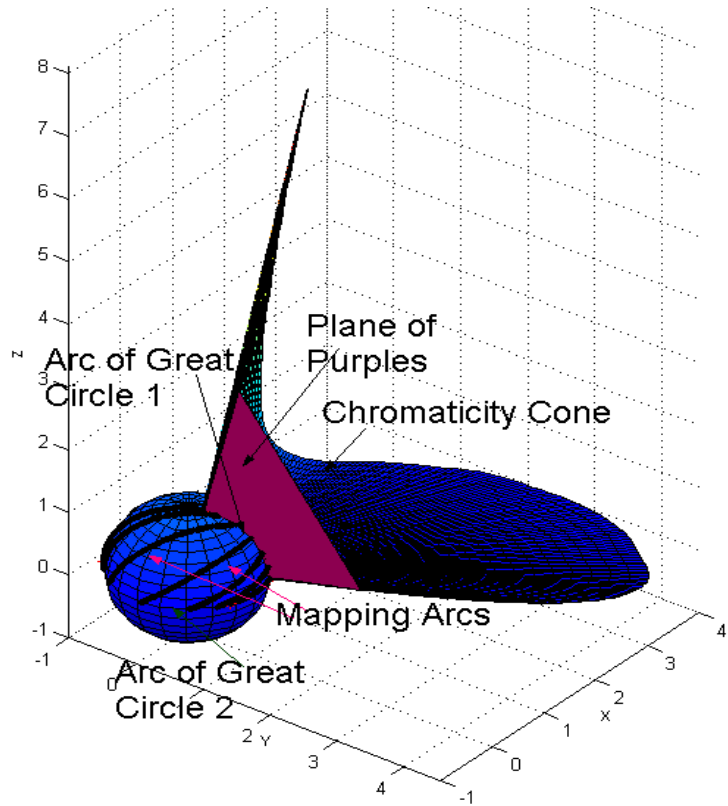

**Fig.11.** To obtain the image of the plane of purples on the chromaticity chart, the arc of great circles of the intersection of the plane of purples with the sphere (called "arc of great circle 1") is mapped bijectively onto the (obtuse) arc of the base circle (also a great circle, called "arc of great circle 2"). The mapping is made through the "mapping arcs" as shown, which are again, arcs of great circles.

4. <u>Metric Induced on the Space Curve (Sphere-CC Intersection)</u>

The analytical condition for the intersection of the sphere and the chromaticity cone is obtained by using the equations of the cone with the sphere equation of radius "r", "h" is the height of the spike:

$$x(\lambda)^2 + y(\lambda)^2 + z(\lambda)^2 = \frac{r^2}{h^2} \tag{3}$$

Taking a total differential at a constant r,

$$(x(\lambda)x(\lambda)'+y(\lambda)y(\lambda)'+z(\lambda)z(\lambda)')d\lambda = -\frac{2r^2}{h^3}dh \qquad (4)$$

The metric induced on the space curve is obtained by eliminating the dh from Eqn (2), Eqn(4). On eliminating, we obtain the metric induced on the space curve of intersection, which is:

$$ds^2 = (-\frac{(x(\lambda)x(\lambda)'+y(\lambda)y(\lambda)'+z(\lambda)z(\lambda)')^2}{(x(\lambda)^2+y(\lambda)^2+z(\lambda)^2)^2} + \frac{(x(\lambda)'^2+y(\lambda)'^2+z(\lambda)'^2)}{(x(\lambda)^2+y(\lambda)^2+z(\lambda)^2)})d\lambda^2 \qquad (5)$$

This is the restriction of the metric of the chromaticity cone to the space curve of intersection (plotted at Fig. 8). On a chosen sphere, the arc length of a chosen point on the space curve is used to locate the corresponding point on the base circle. From the construction, it is clear that the correspondence is a bijection.

5. The isoX, isoY (iso-luminance) and isoZ Lines on the CC

The CC is sectioned with planes of the following type: X=constant, Y=constant, Z=constant. These plane curves are then mapped on to the chromaticity chart and plotted using a color code: isoX is marked in red, isoY (iso-luminance) are marked in green and isoZ are marked in blue.

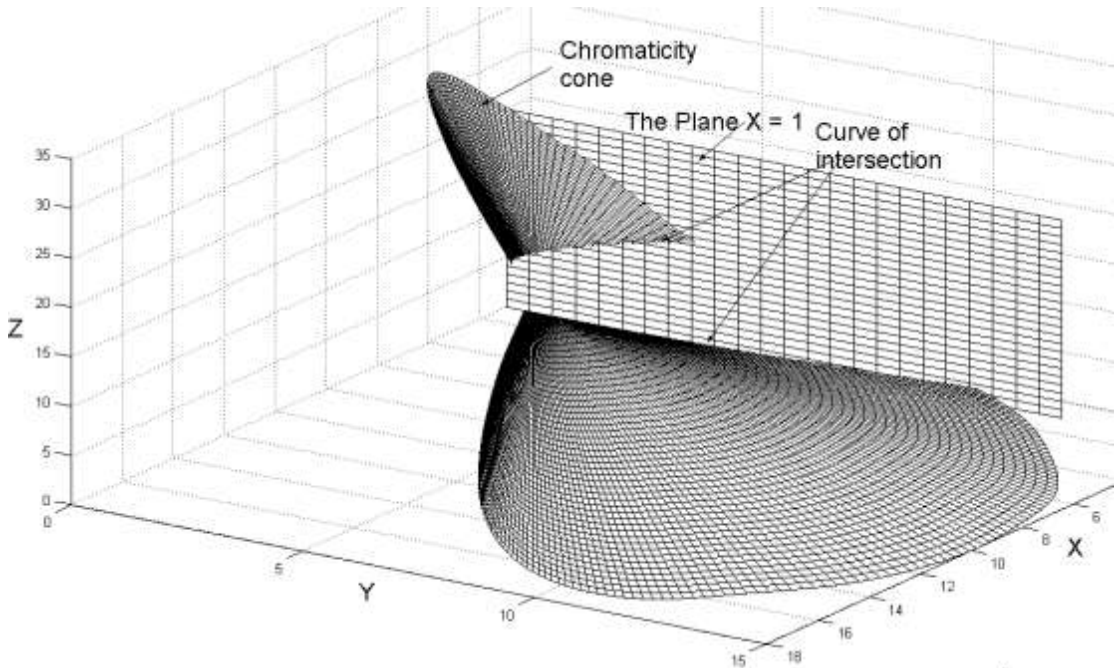

**Fig.12.** The isoX curves on the chromaticity chart are the intersections of the X=constant planes with the chromaticity cone. The curve of intersection corresponding to the X=1 plane is marked in the figure. Sometimes, the X=constant intersections are elongated curves on the chromaticity chart. Often because of the chosen dimensions for view, only two branches of the elongated curve are visible.

The isoX line on the Z=1 plane is at fig and the corresponding lines

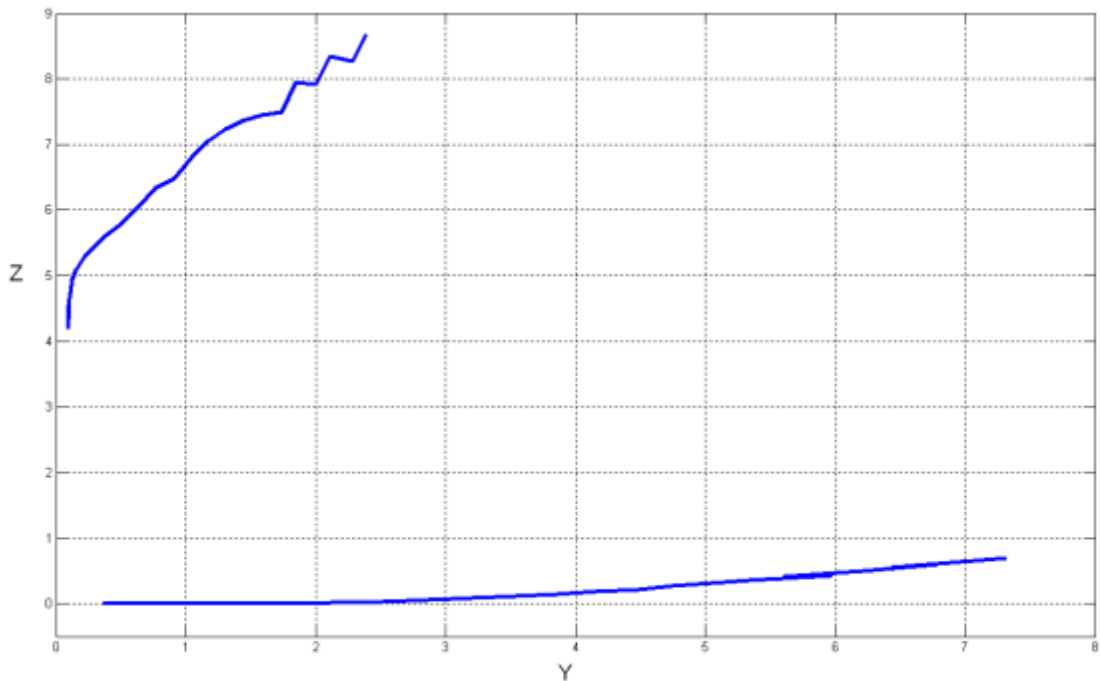

**Fig.13.** The isoX curve for X=1 is plotted on the YZ plane. Two branches of the curve are visible.

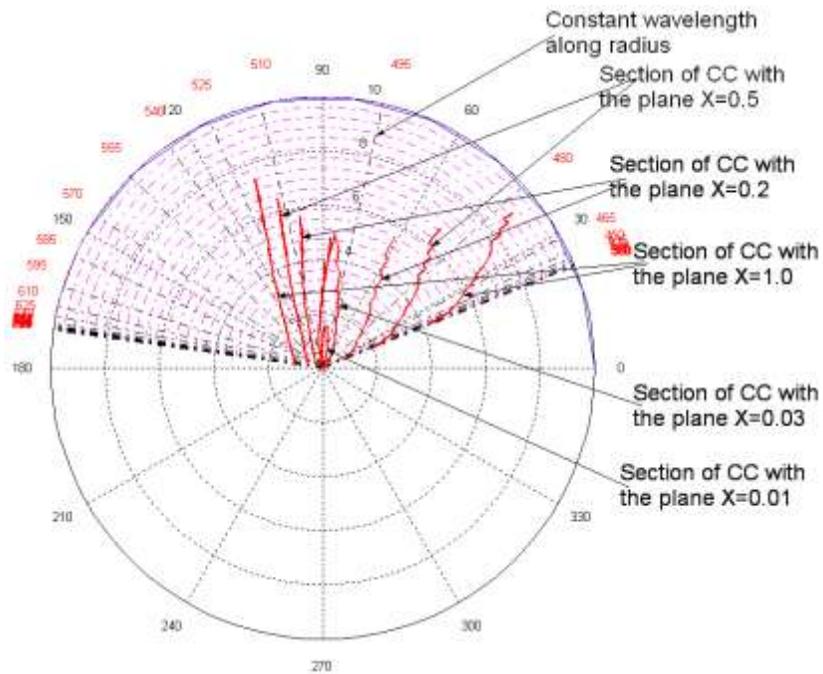

**Fig.14.** Several isoX lines (corresponding to X=0.01, 0.03, 0.2, 0.5, 1.0) are plotted on the chromaticity chart. The closed loops of the isoX lines corresponding to X=0.01, 0.03 are visible. The other loops are too long to fit in the chosen dimensions of the chart, so only two branches of each of those loops are visible. The isoX curves are all shown in red.

Likewise, the isoY lines are constructed by sectioning the CC with Y=1 plane and then mapping the section on the chromaticity chart.

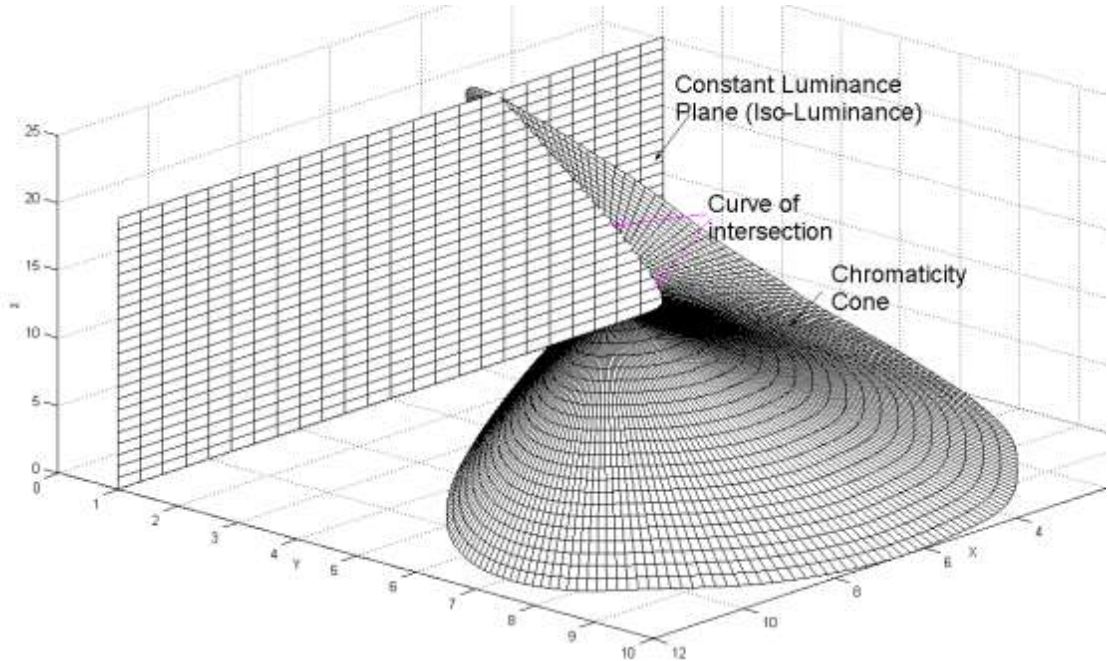

**Fig.15.** Iso-Luminance lines (or isoY lines) on the chromaticity chart are the images of the intersection of the constant luminance planes (Y=constant) with the chromaticity cone. The (plane) curve of intersection is marked in the figure.

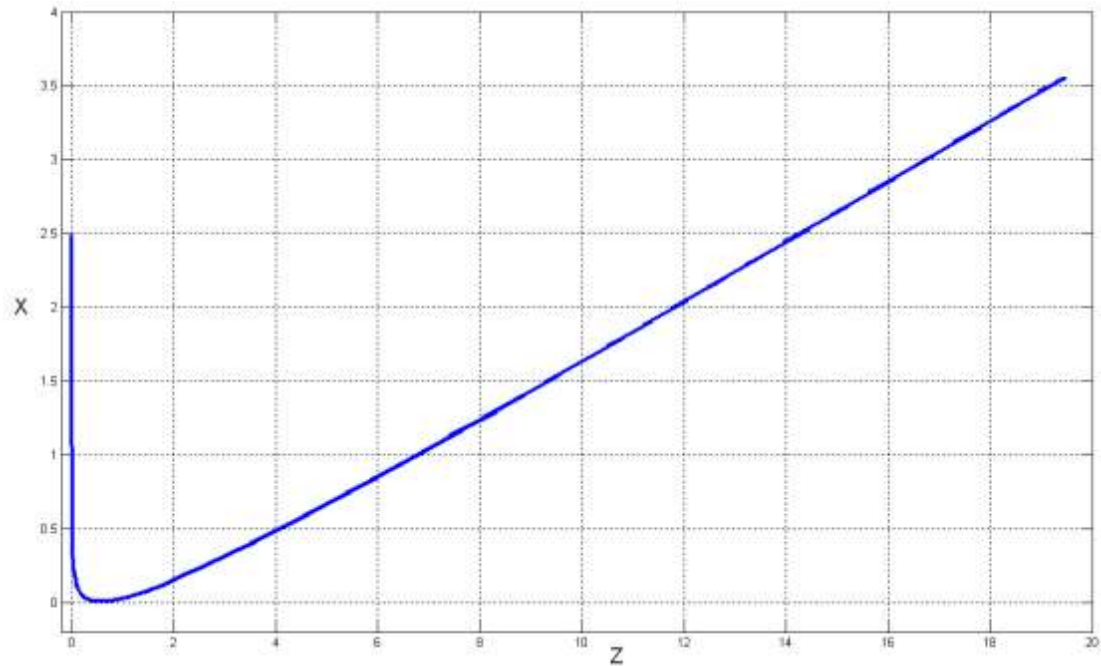

**Fig.16.** The iso-luminance curve (isoY curve) on the Y=1 plane is plotted.

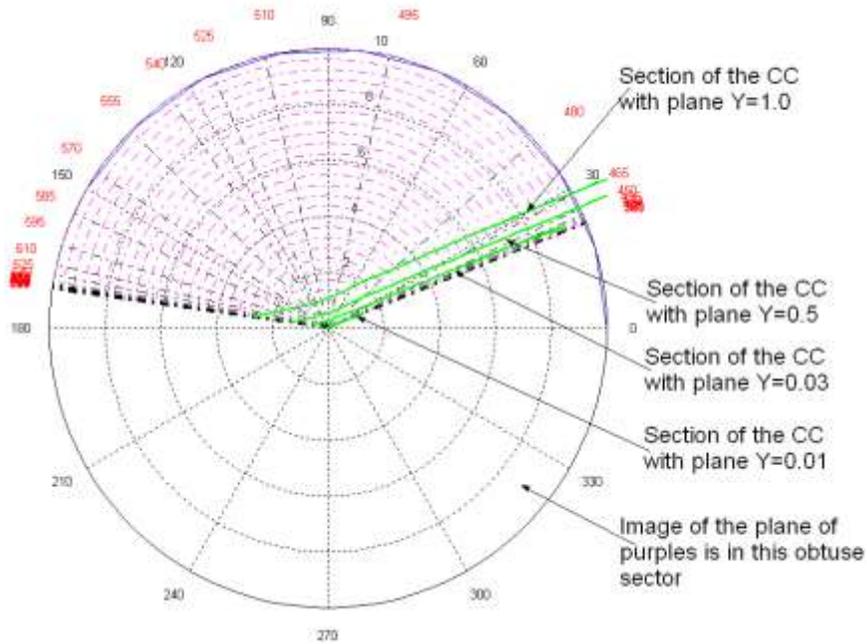

**Fig.17.** IsoY or iso-luminance lines corresponding to Y=0.01, 0.03, 0.5, 1.0 are shown. These are the constant luminance curves on the chromaticity chart. The curves are plotted in green.

The isoZ lines are obtained by sectioning the CC with Z=1 plane and plotting the section on the chromaticity chart.

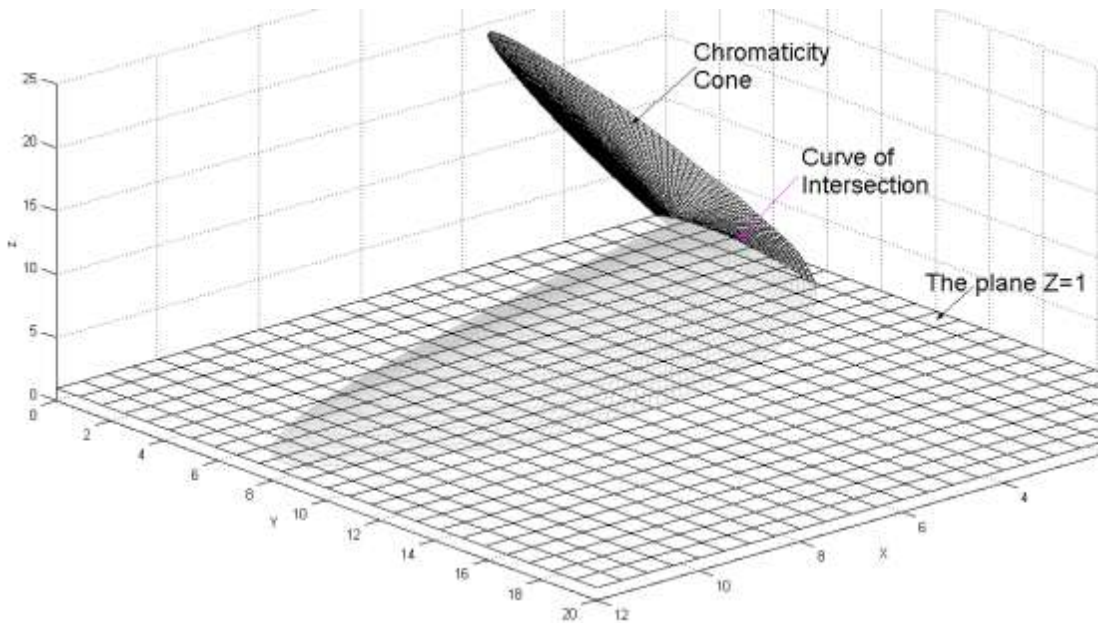

**Fig.18.** The isoZ curves on the chromaticity chart are the intersections of the Z=constant planes with the chromaticity cone. The curve of intersection corresponding to the Z=1 plane is marked in the figure.

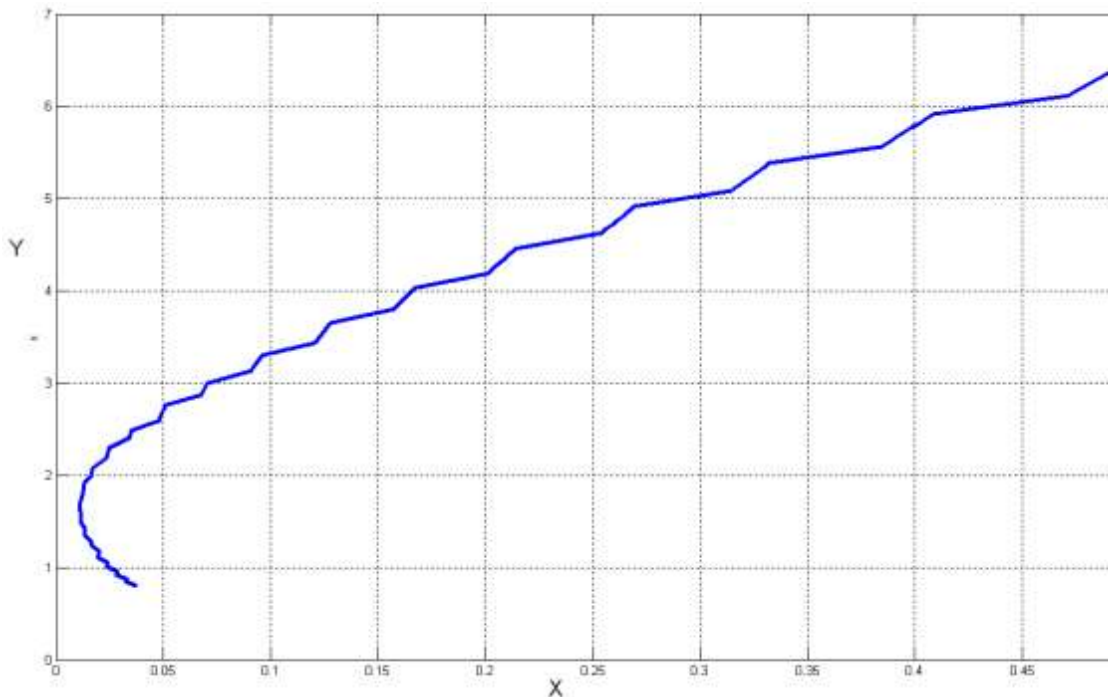

**Fig.19.** The isoZ curve for Z=1 is plotted on the XY plane. The feasible target error for the solution algorithm caused some undulations on the solution curve.

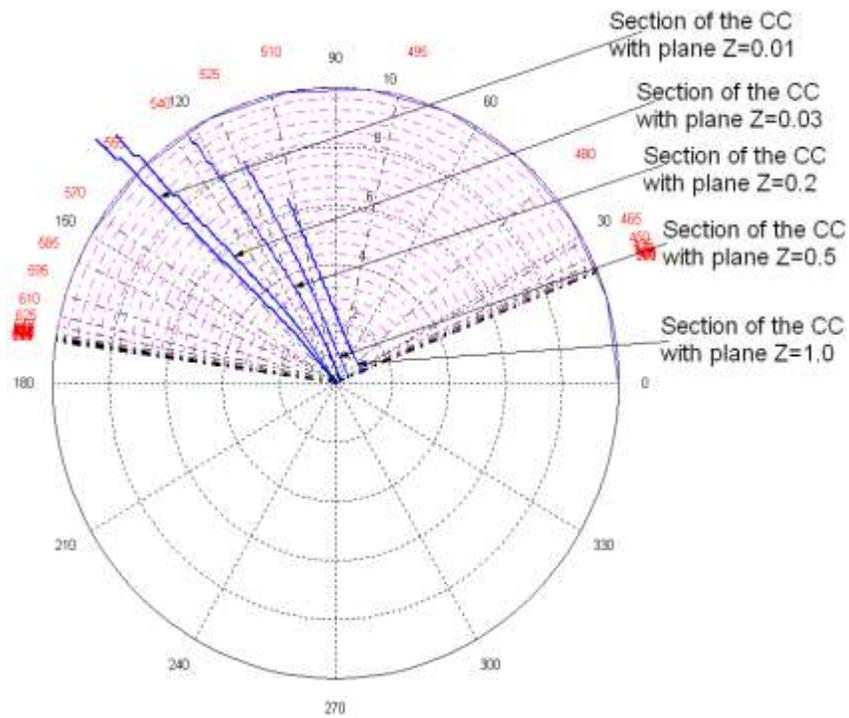

**Fig.20.** The isoZ lines corresponding to Z=0.01, 0.03, 0.2, 0.5, 1.0 are plotted in blue on the chromaticity chart.

The three types of sectional curves are now superposed on a single chromaticity chart keeping to the color code:

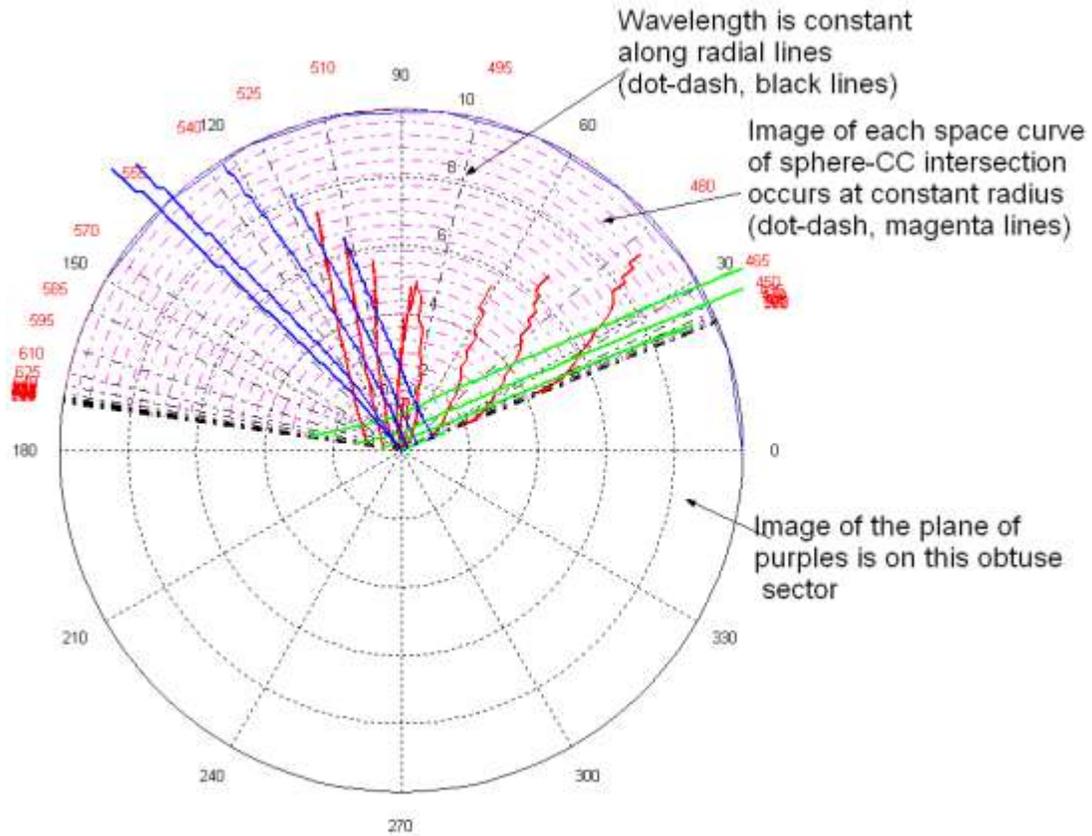

**Fig.21.** The isoX, isoY (iso-luminance lines) and isoZ are all superposed onto the chromaticity chart, using the same color code used in the previous diagrams. The relative positions and orientations of these curves is now evident.

Salient features of the chromaticity chart are:
1. The wavelengths are constant along the radial lines, marked in dot dashed black lines.
2. Images of the sphere-cc intersection occur at constant radii and marked in dot-dashed magenta lines.
3. The obtuse sector of 210 degrees is the image of the plane of purples.
4. The isoX, isoY and isoZ lines are marked.

6. <u>Spectral Locus on the Chromaticity Chart</u>

The spectral locus is the intersection of the CC with the normalizing plane, X+Y+Z=1. These equations are solved simultaneously and the solution data is mapped onto the chromaticity chart.

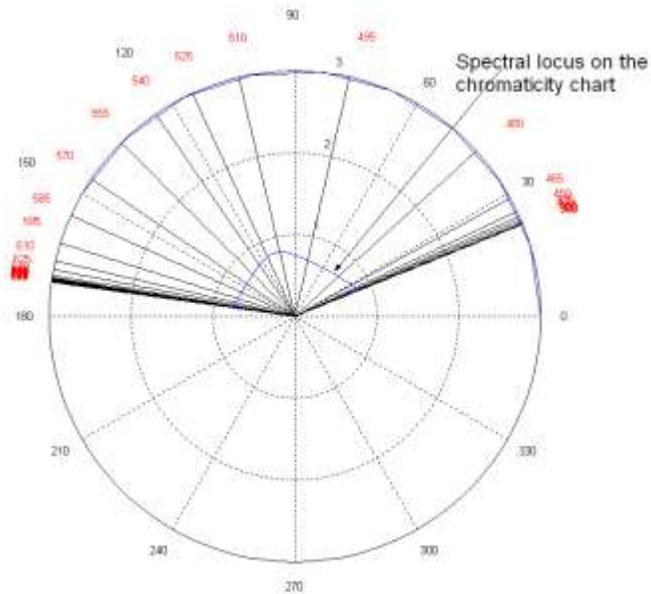

**Fig.22.** The image of the spectral locus on the chromaticity chart. This is the solution of the normalizing plane (X+Y+Z=1) with the chromaticity cone.

7. Compactification of the Chromaticity Chart

We add infinity to the chromaticity chart and compactify the entire chart plus infinity. The stereographic projection is then used to map the chart plus infinity onto a compact sphere.

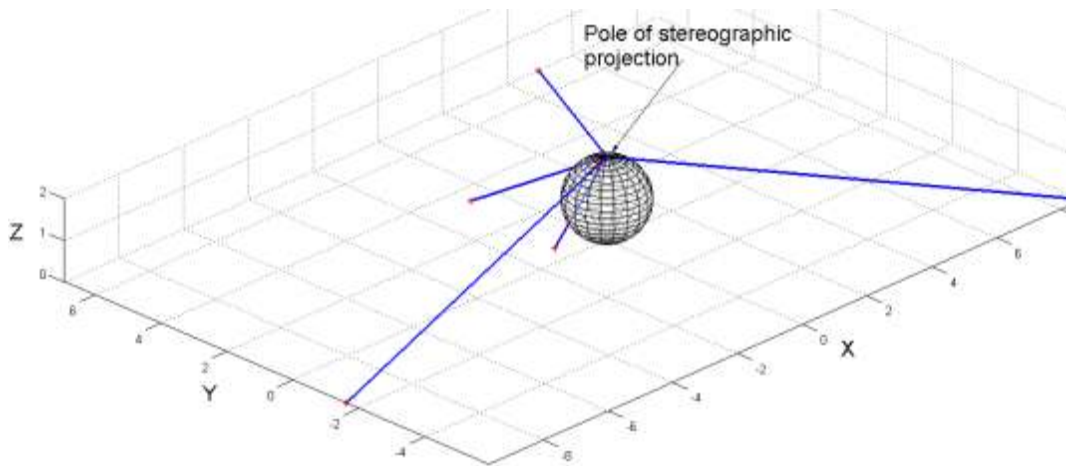

**Fig.23.** The stereographic projection maps the entire plane plus infinity onto the compact sphere. As shown, straight line from the pole of the sphere connecting an arbitrary point on the plane, maps that point to the point of intersection of the line with the sphere. The infinity gets mapped to the pole.

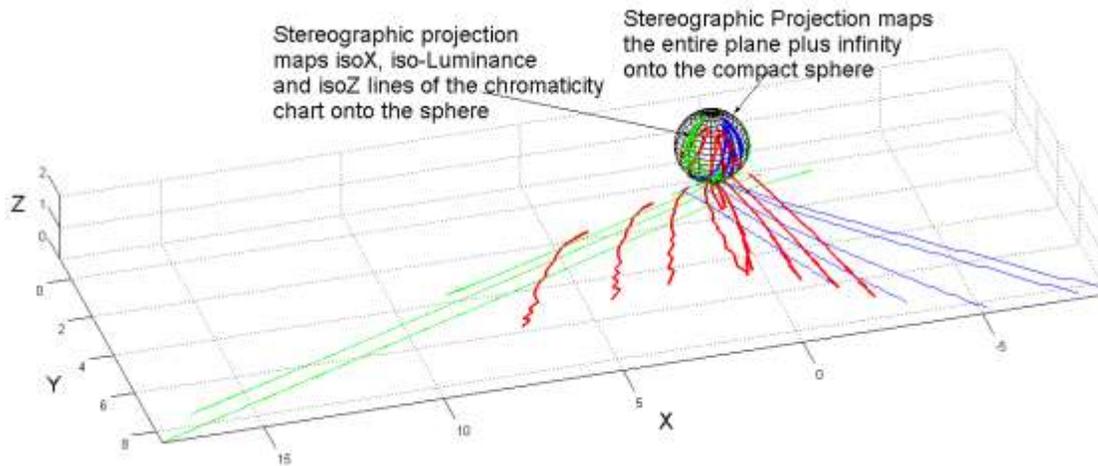

**Fig.24.** On adding infinity to the chromaticity chart, a compact representation of the entire chart plus infinity is obtained. The stereographic projection then maps the chart plus infinity onto a compact sphere. The north pole of the sphere is used as the pole for the stereographic projection.

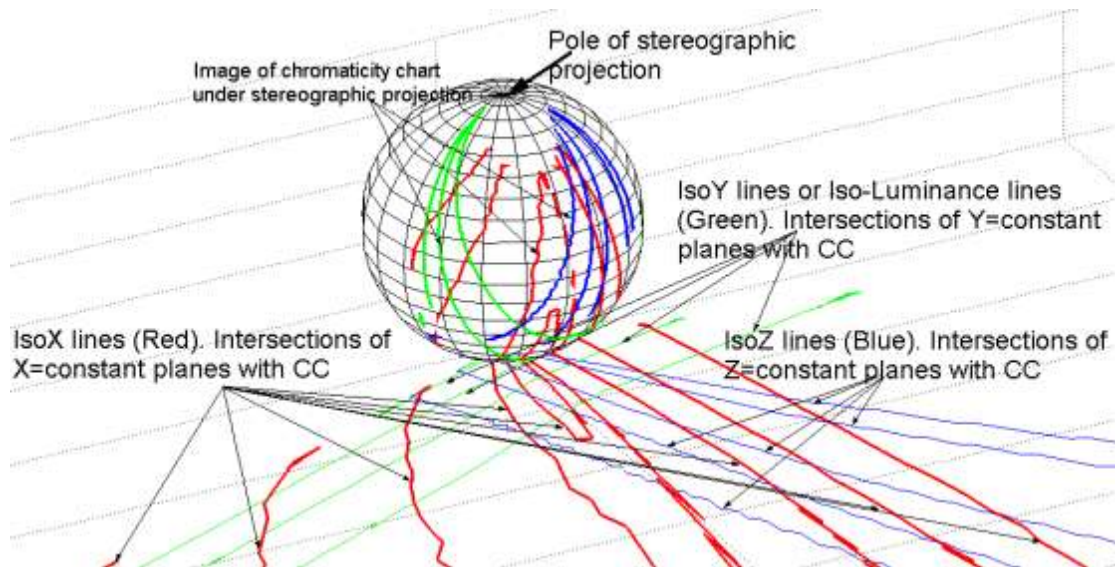

**Fig.25.** A magnification of the previous diagram shows the details of the construction and the stereographic projection.

## 8. Convexity of the Chromaticity Cone

In the XYZ space the chromaticity cone is convex. Therefore this construction is sufficient for practical purposes. An arbitrary linear transformation on the color matching functions could cause the corresponding chromaticity cone to go non-convex. However, the chromaticity chart may still be constructed using similar methods described.

## 9. Some Examples

Some examples are below. The figures are self-explanatory.

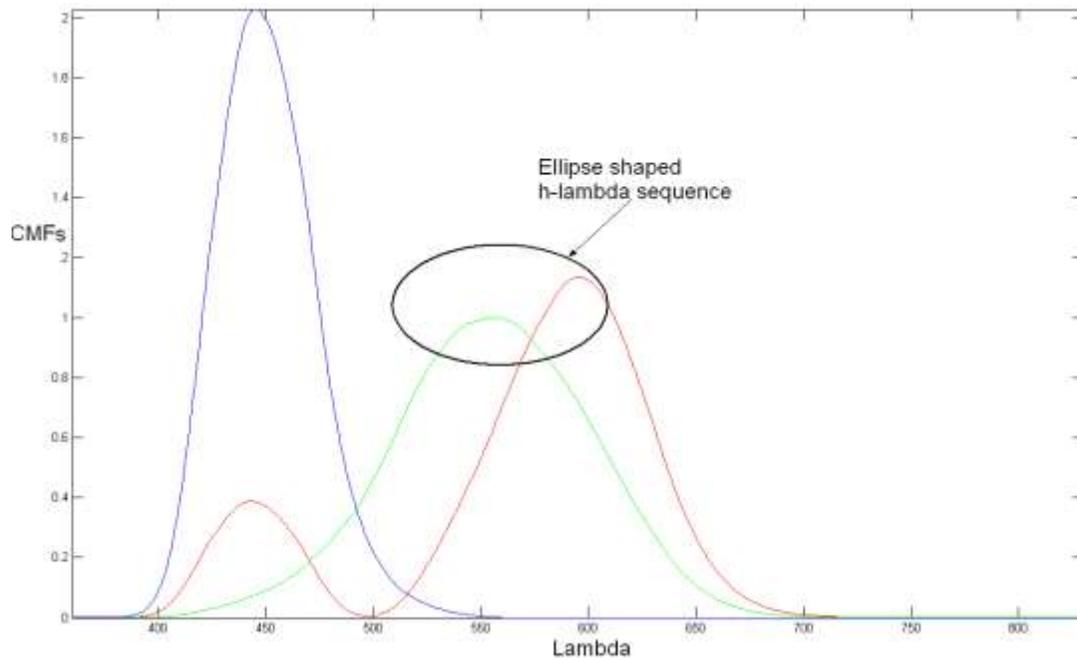

**Fig.26.** An ellipse type h-lambda sequence is chosen and plotted on the color matching functions diagram. This ellipse is a spike-type time ordered spectral radiance distribution. The image of this ellipse is plotted on the chromaticity cone and the chart. (For the spike 'h', tracing the shown ellipse, the ordinate is spectral radiance, assume suitable orientation – say clockwise)

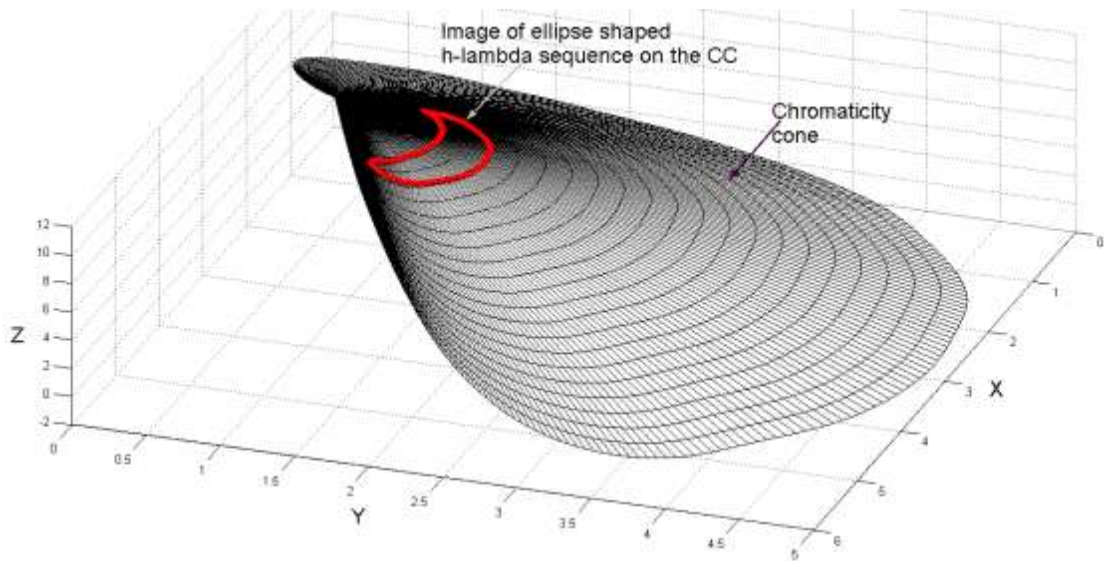

**Fig.27.** The image of the ellipse shaped SRD is shown plotted in red, on the chromaticity cone.

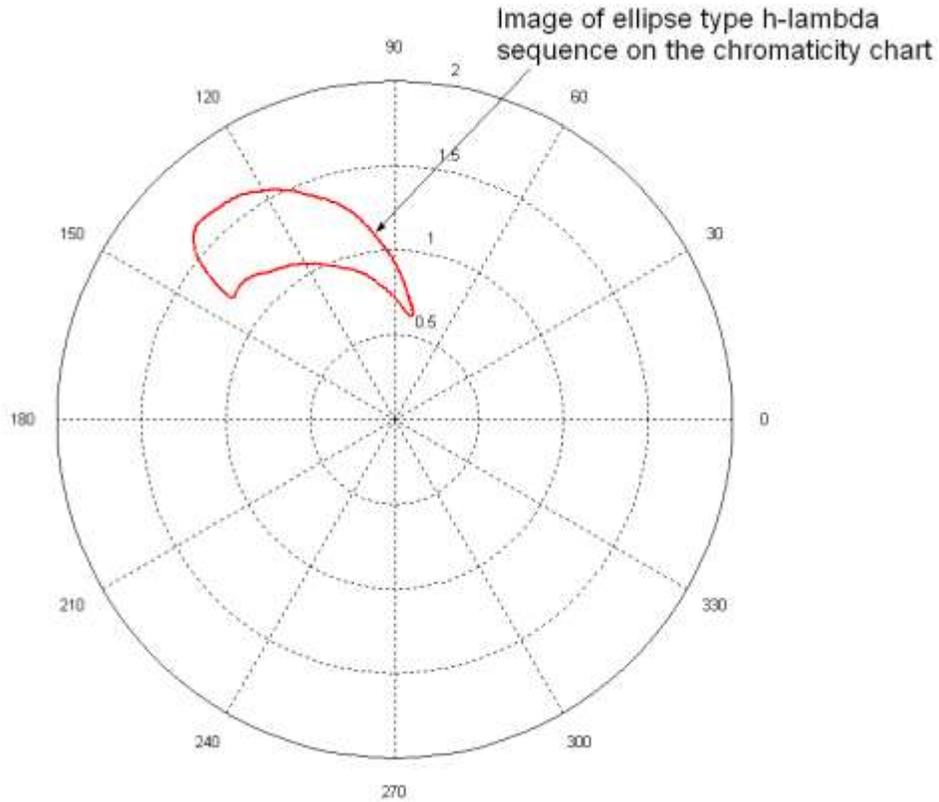

**Fig.28.** The image of the ellipse type h-lambda spike sequence is now plotted on the chromaticity chart.

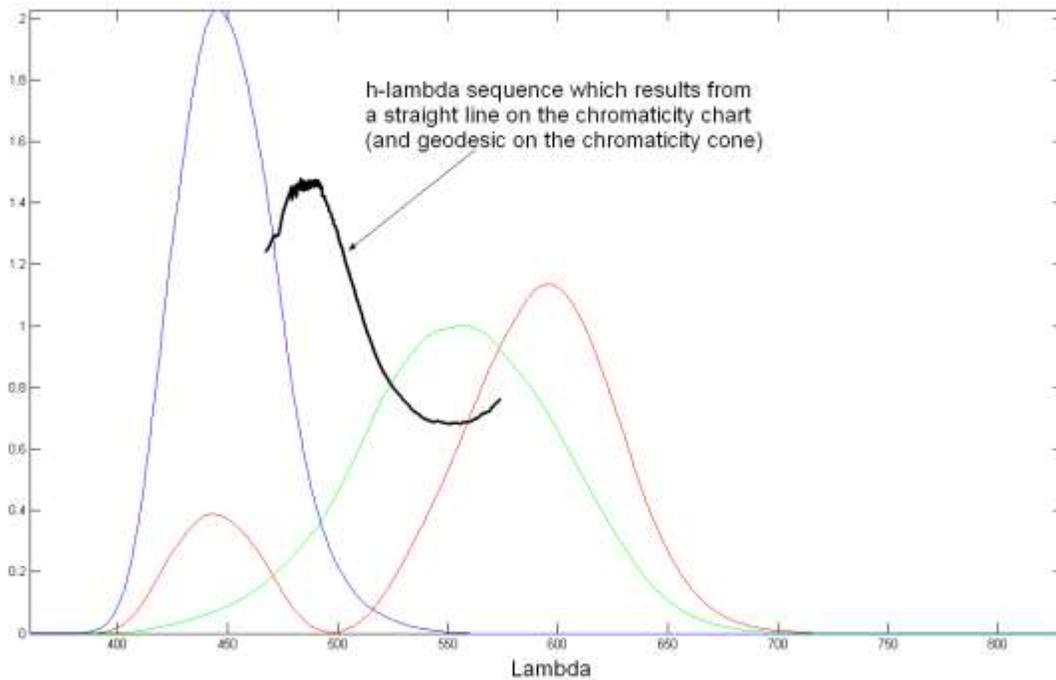

**Fig.29.** By inverting the straight line (geodesic of CC) of the chromaticity chart, we obtain the h-lambda (spike-type) sequence of SRD.

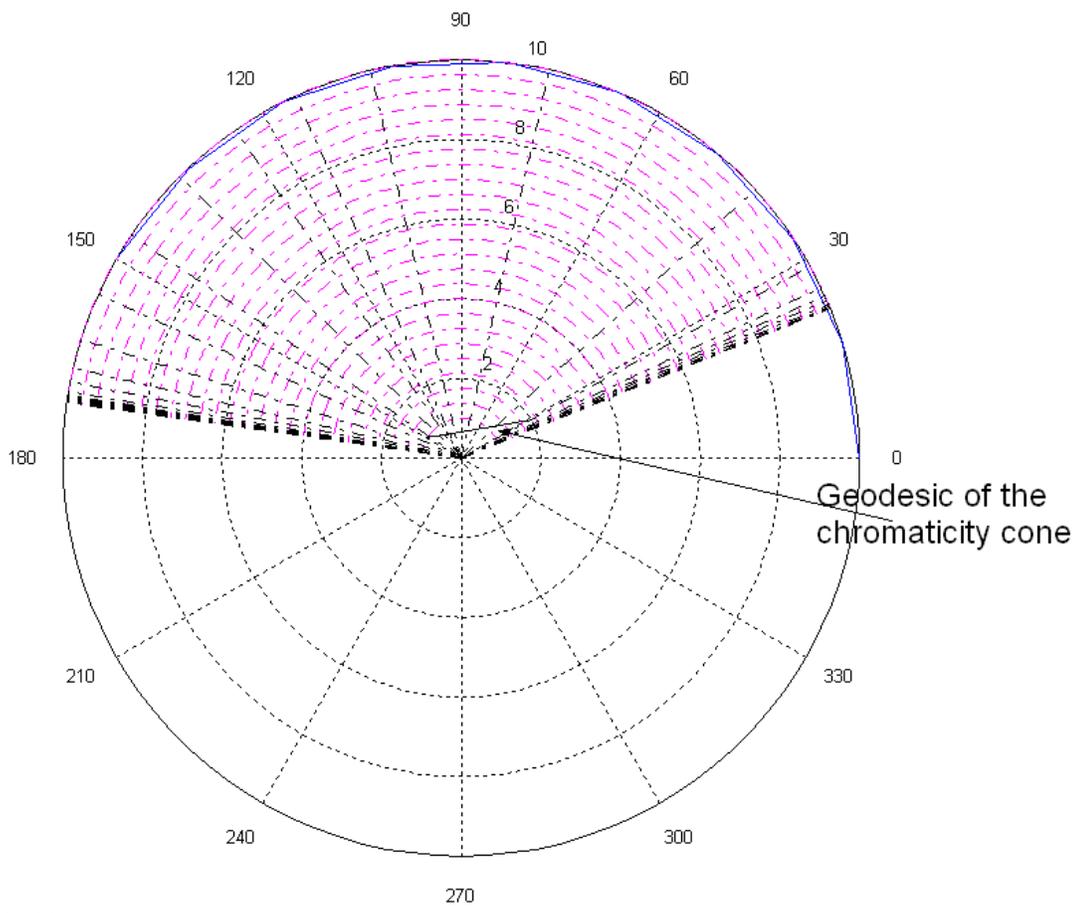

**Fig.30.** Because the chromaticity cone (in the XYZ space) is convex surface, the straight lines of the chromaticity chart are the geodesics of the cone.

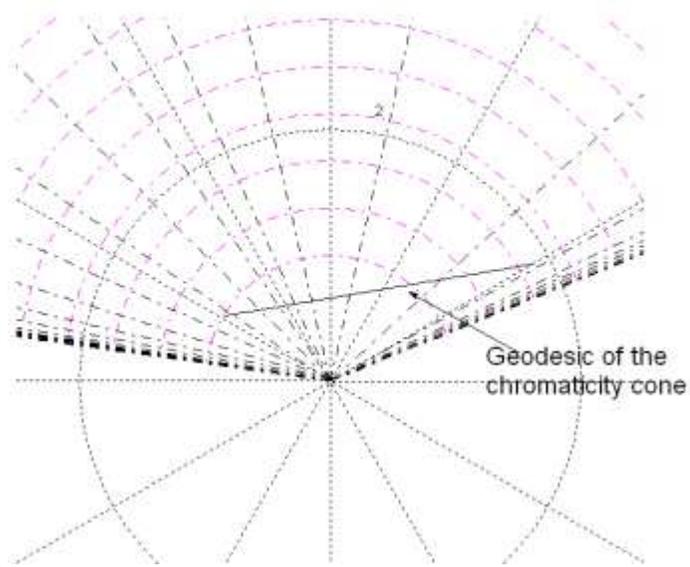

**Fig.31.** A magnification of the geodesic shown in the previous diagram.

Conclusion:
The chromaticity cone is opened up as a flat chart and the loci corresponding to constant X, Y (iso-luminance) and Z are plotted. Geodesics of the chromaticity cone are also plotted. In future articles, the construction herein described would be used to formulate new properties of color perception. In future articles, the present construction would be enhanced and new designs for color perception would be developed.

[*]Acknowledgment
Author was formerly at the Laboratory for Perceptual Dynamics, RIKEN Brain Science Institute, Saitama, Japan. For the contents of this article early verifiable dates are available.